\documentclass[3p]{elsarticle}
\usepackage[utf8]{inputenc}
\usepackage[english]{babel}
\usepackage{amsmath}
\usepackage{amsfonts}
\usepackage{amssymb}
\usepackage{mathrsfs}
\usepackage{makeidx}
\usepackage{graphicx}   
\usepackage{siunitx}
\usepackage{float}
\usepackage{hyperref}
\hypersetup{
    colorlinks=true,
    linkcolor=blue
    }
\usepackage{url}
\usepackage{caption} 
\usepackage{array}
\usepackage{wasysym}
\usepackage{textcomp}
\usepackage{setspace}
\makeatletter
\usepackage{fancyhdr}
\usepackage{titlesec}
\usepackage[flushleft]{threeparttable}

\title{A Titan mission using
the Direct Fusion Drive}
\author[1,2]{Marco Gajeri}
\ead{marco.gajeri@gmail.com}
\author[1,2]{Paolo Aime}
\ead{paolo.aime.95@gmail.com}
\author[2,3,4]{Roman Ya. Kezerashvili}
\ead{RKezerashvili@citytech.cuny.edu}
\address[1]{Politecnico di Torino, Torino, Italy}
\address[2]{New York City College of Technology, The City University of New York, New York, USA}
\address[3]{The Graduate School and University Center, The City University of New York, New York, USA}
\address[4]{Samara National Research University, Samara, Russian Federation}

\date{September,26 2020}

\begin{document}
\begin{abstract}
The main purpose of this work is to perform an analysis of realistic new trajectories for a robotic mission to Saturn's largest moon, Titan, in order to demonstrate the great advantages related to the Direct Fusion Drive (DFD). The DFD is a D -$^3$He fuelled, aneutronic, thermonuclear fusion propulsion system, related to the ongoing fusion research at Princeton Plasma Physics Laboratory (PPPL) \cite{cohen2017direct}. This fusion propulsion concept is based on a magnetically confined field reversed configuration plasma, where the deuterium propellant is heated by fusion products, and then expanded into a magnetic nozzle, providing both thrust and electrical energy to the spacecraft. The trajectories calculations and analysis for the Titan mission are obtained based on the characteristics provided by the PPPL \cite{cohen2017direct}. Two different profile missions are considered: the first one is a thrust-coast-thrust profile with constant thrust and specific impulse; the second scenario is a continuous and constant thrust profile mission, with a switch in thrust direction operated in the last phases. Each mission study is divided into four different phases, starting from the initial low Earth orbit departure, the interplanetary trajectory, Saturn orbit insertion and the Titan orbit insertion. For all mission phases, maneuver time and propellant consumption are calculated. The results of calculations and mission analysis offer a complete overview of the advantages in term of payload mass and travel time. 
It is important to emphasize that the deceleration capability is one of the DFD game changer: in fact, the DFD performance allows to rapidly reach high velocities and decelerate in even shorter time period. This capability results in a total trip duration of 2.6 years for the thrust-coast-thrust profile and less than 2 years considering the continuous thrust profile. The high payload enabling capability, combined with the huge electrical power available from the fusion reactor, leads to a tremendous advantage compared to present technology.
\end{abstract}

\maketitle

\section{Introduction} 
The emotional desire to explore and challenge the boundaries of our knowledge has allowed us to evolve and make amazing discoveries. As our ancestors searched and discovered new lands so we are exploring the universe, looking for pleasant answers. Because of its relative proximity, human exploration began with Earth's natural satellite, taking first men to walk on lunar soil in 1969 \cite{godwin1999apollo}. At the beginnings of the sixties first robotic Mars missions were designed and launched \cite{launius1997chronology, branigan1965mariner, chappell1995mars} focusing to the payload mission capability and mission time. The higher the payload mass, the more scientific instruments can be carried on-board the spacecraft and the more precious scientific data are collectable. This is very important both for robotic and manned missions, such as those planned for the near future for the Moon and Mars. It would be significant to increase the payload as much as possible, without excessively extend the journey time. One can say that new propulsion concept need to be developed in order to colonize our Solar system, overcoming the limitations related to chemical and electric propulsion (CP and EP respectively). In fact, considering current solutions, low power EP systems are affected by long journey time because of the extremely low thrust, even though their high specific impulse $I_{sp}$ ($ \approx 1500 - 5000$ s). CP systems are not convenient due to the limitation on the maximum specific impulse ($I_{sp} \approx 450 $ s) \cite{arbit1970investigation}, directly dependent on the fuel chemical energy.

The main issue for human space exploration is the huge vastness of space, that condemns space travellers to lengthy travel times, forcing the crew to face many problems such as radiation exposure and microgravity conditions. In a long journey in space physiological response to microgravity adaptation has all the features of accelerated ageing involving almost all body systems \cite{aubert2005cardiovascular}. Surely, before humanity can succeed in a solar system manned mission, a great technological advancement will have to be carried out, which must concern many fields with a particular focus on propulsion. At present, taking for example a Mars mission, round-trip may take up to three years and research suggests that astronauts could lose close to half their bone mass before they return \cite{samarabandu1993analysis}. Therefore, it is natural to think about using nuclear energy to propel spacecraft which could reach Mars in about half the time of current missions. Using the the Direct Fusion Drive (DFD) \cite{cohen2017direct} one way trips to Mars in slightly more than 100 days become
possible \cite{genta2020achieving}, while the total mission duration for exploration of trans-Neptunian objects takes slightly more than eight years \cite{AimeGajeri2020}.  

The exploration of the solar system requires advanced propulsion techniques capable of specific impulse above $10^4$ s and specific power in the range 1-10 kW/kg \cite{romanelli2005assessment}. The idea of using nuclear power for spacecraft propulsion arises from the high energy density of the fuel and the high velocity of the fusion products, which resulted into the establishment of the NERVA project by NASA in 1960 \cite{robbins1991historical}. Even today, the research activities related to both fission and fusion propulsion systems are in development and they will be able to overcome the limitations tied to classical propulsion systems. 

Our work focuses on the great advantages related to the Direct Fusion Drive project, that would enable faster deep space missions. The paper is organized in the following way: in Section \ref{DFD} is given a brief description of the DFD engine, its main characteristics, fuel choice, and the thrust model. 
The input characteristics for Earth - Titan mission are presented in Section \ref{mission}.
Two different Titan mission profiles are analysed and discussed in Section \ref{tctmission} and \ref{ct}: i. the thrust-coast-thrust profile; ii. the continuous thrust profile. Section \ref{confronto} highlights the differences between the two analysis performed, with particular focus on the mission times, propellant consumption and payload. The conclusions follow in Section \ref{conclusioni}.

\section{Direct Fusion Drive} \label{DFD}
The Direct Fusion Drive is a revolutionary fusion propulsion concept that would produce both propulsion and electric power from a single, compact fusion reactor \cite{razin2014direct}. The project, funded by NASA, 
is based on the overwhelming advantages offered by the ongoing Princeton field reversed configuration (PFRC-2) fusion experiment at Princeton Plasma Physics Laboratory (PPPL) \cite{cohen2011rf}. The purpose of PPPL research is to find solutions for the critical scientific and technological problems related to fusion technology. DFD concept suits to several kind of space destinations, such as Mars manned and robotic missions, heavy cargo missions to the outer solar system or the near interstellar space \cite{cohen2017direct, thomas2017fusion, genta2020achieving, AimeGajeri2020}.

\subsection{Princeton field reversed configuration}
The PFRC-2 concept employs a unique radio frequency (RF) plasma heating method, known as “\textit{odd-parity} heating”, which increases the plasma temperature in order to achieve the proper physics conditions which enable the fusion process in a field reversed configuration (FRC) plasma \cite{cohen2000ion, cohen2007stochastic}. The FRC is a particular magnetic field geometry, accidentally discovered in the sixties \cite{kolb1959field}, in which a toroidal electric current is induced inside a cylindrical plasma, creating a poloidal magnetic field. The latter is reversed with respect to the direction of an externally applied axial magnetic field. This new heating method, invented by Cohen and Milroy \cite{cohen2000maintaining}, is based on a magnetic field that is antisymmetric about the mid-plane normal to the axis and added to a FRC plasma maintaining its closed field line structure. It was first theorized in 2000 \cite{cohen2000maintaining} and demonstrated in 2006 (PFRC-1) \cite{cohen2007formation}. This is a crucial point, because the “open” field lines let the plasma to escape and consequently reduce confinement time, which is tightly bound to optimal fusion conditions \cite{guo2005observations}. The fusion process is magnetically confined in the core, the region inside the magnetic \textit{separatrix}, 
which is an imaginary closed surface that demarcates “open” magnetic field lines, those that cross the device walls, from those that stay fully inside the device. The “open” field line region - also called the scrape-off layer (SOL) - is the region where the cold deuterium propellant is heated by the fusion products.

The core needs a strong plasma current perpendicular to the FRC’s magnetic field to form the closed magnetic-field lines. Otherwise, the configuration will have instability problems and it will destroy itself \cite{rosenbluth1979mhd}. More specifically, PFRC exploits a rotating magnetic field (RMF$_o$) with \textit{odd-parity} symmetry, produced by the oscillation of the current in four quadrature-phased RF antennae \cite{cohen2017direct}. Two pairs operate 90 degrees out of phase on adjacent sides of the plasma and generate RMF$_o$ which is about $0.1-5$\% the strength of the axial magnetic field. Then, the magnetic field on one side of each 8-shaped antenna has a direction opposite to the other side and closed field lines in the generated FRC keep the plasma trapped when it is heated. Therefore, a toroidal current is induced by RMF$_o$ in the plasma confined by the externally-applied axial magnetic field. Then, this current induces a poloidal closed magnetic field, which improves the plasma confinement. 
Therefore, $RMF_o$ generates the current and heats the plasma ions and electrons \cite{glasser2002ion}, leading to compact devices with excellent stability \cite{welch2010formation} due to the fact that a small, high-temperature FRC plasma, has less problems against instability than other fusion devices. See Ref. \cite{steinhauer2011review, ishida1988variational} for more details.

\subsection{Fuel choice and neutron production}
Fusion of light nuclei produces much more energy per unit mass than fission processes. Usually one of the components of the fusion reaction is proton, deuterium or tritium. The other component involved into the fusion of light nuclei can be another deuterium, isotopes of helium, $_{2}^{3}$He or $ _{2}^{4}$He, and isotopes of lithium, $_{3}^{6}$Li and $_{7}^{7}$Li. The region where fusion reactions take place in DFD is the high temperature, moderate density plasma region named the core. The fusion reaction of nuclei of deuterium (D) and tritium (T) is the most promising for the implementation of controlled thermonuclear fusion, since its cross section even at low energies is sufficiently large \cite{dolan2013magnetic}. However, due to significant emission of neutron the D$-$T fuel is not the best choice for the DFD and aneutronic fuel such as the mixture of the deuterium and helium-3 isotope, D$-$ $_{2}^{3}$He, is most preferable. The choice of D$-$ $_{2}^{3}$He fuel mixture to produce the D--$_{2}^{3}$He plasma is related to the neutrons production problem. If neutrons are produced from the fusion reactions, a certain amount of energy is not usable, leading to not negligible losses of energy, as well as the contamination of a spacecraft due to the neutrons emission, which should be shielded to protect the spacecraft and crew. Neutrons are hard to “direct” due to the fact that they have no charge and can not be controlled with electric or magnetic fields.
Therefore, it is essential to reduce the neutron fluxes in order to minimize the damage and activation of nearby materials and structures and
consequently the shielding mass.

Let us focus on the fusion processes in D--$_{2}^{3}$He plasma. Depending on
plasma temperature the ignition of D--$_{2}^{3}$He, D$-$D, and $_{2}^{3}$%
He--$_{2}^{3}$He fusion can occur. Therefore the D--$_{2}^{3}$He plasma can
admit the following aneutronic and neutron emitted primary reactions: 
\begin{eqnarray}
\text{D}+\text{ }_{2}^{3}\text{He} &\rightarrow &\text{ }_{2}^{4}\text{He}+p%
\text{ \ \ \ \ \ \ \ }(Q=18.34\text{ MeV),}  \label{DD3He} \\
&\rightarrow &\text{ T}+2p  \label{DD3He1} \\
&\rightarrow &\text{ D}+\text{D}+p,  \label{DD3He2} \\
&\rightarrow &\text{ }_{3}^{5}\text{Li }+\gamma ;  \label{DD3He3} \\
\text{D}+\text{D} &\rightarrow &\text{ }_{2}^{3}\text{He}+n\text{ \ \ \ \ \
\ \ }(Q=3.25\text{ MeV}),  \label{DD} \\
&\rightarrow &\text{ T}+p\text{\ \ \ \ \ \ \ \ \ \ \ }(Q=4.04\text{ MeV);}
\label{DD1} \\
_{2}^{3}\text{He}+\text{ }_{2}^{3}\text{He} &\rightarrow &\text{ }_{2}^{4}%
\text{He}+2p\text{ \ \ \ \ \ \ }(Q=12.86\text{ MeV),}  \label{He3He3} \\
&\rightarrow &\text{ D}+\text{D}+2p,  \label{He3He31} \\
&\rightarrow &\text{ }_{3}^{5}\text{Li }+p,  \label{He3He32} \\
&\rightarrow &\text{ }_{3}^{5}\text{Li}^{\ast }\text{ }+p,  \label{He3He33}
\\
&\rightarrow &\text{ }_{4}^{6}\text{Be}+\gamma .  \label{He3He34}
\end{eqnarray}%
In reactions (\ref{DD3He}) $-$ (\ref{He3He34}) D ($_{1}^{2}$H) and T ($%
_{1}^{3}$H) are notation for the isotopes of the hydrogen -- deuterium and
tritium nuclei and the energy liberated ($Q$ value)\ for reactions (\ref%
{DD3He}), (\ref{DD})$-$(\ref{He3He3}) are given in parentheses. All primary reactions in D$-$ $^{3}$He are aneutronic accept the process (\ref{DD}), where 2.45 MeV energy neutrons are produced. The D$-$D has two almost equally exothermic channel (\ref{DD}) and (\ref{DD1}). The ignition of D$-$D fusion requires the plasma temperature about 5$\times 10^{8}$ K. Due to production of tritium (\ref{DD1}) in D$-$D fusion, D$-$ $^{3}$He plasma admits the secondary fusion processes: deuterium--tritium, tritium--tritium
and tritium--helium-3. These secondary processes have the following branches for nuclear reactions:

\begin{eqnarray}
\text{D}+\text{T} &\rightarrow &\text{ }_{2}^{4}\text{He}+n\text{ \ \ \ \ \
\ \ \ \ \ \ }(Q=17.6\text{ MeV),}  \label{DT} \\
&\rightarrow &\text{ \ D}+\text{D}+n,  \label{DT2} \\
&\rightarrow &\text{ \ }_{2}^{3}\text{He}+2n;  \label{DT3} \\
\text{T}+\text{T} &\rightarrow &\text{ }_{2}^{4}\text{He}+2n\text{ \ \ \ \ \
\ \ \ \ \ }(Q=11.3\text{ MeV),}  \label{TT} \\
&\rightarrow &\text{ \ D}+\text{D}+2n,  \label{TT1} \\
&\rightarrow &\text{ }_{2}^{5}\text{He}+n;  \label{TT2} \\
\text{T}+\text{ }_{2}^{3}\text{He} &\rightarrow &\text{ }_{2}^{4}\text{He }+%
\text{D \ \ \ \ \ \ \ \ \ \ }(Q=14.3\text{ MeV),}  \label{THe3} \\
&\rightarrow &\text{ }_{2}^{4}\text{He}+n+p\text{ \ \ \ \ \ \ }(Q=12.1\text{
MeV),}  \label{THe32} \\
&\rightarrow &\text{ }_{2}^{5}\text{He}+p,  \label{THe33} \\
&\rightarrow &\text{ }_{2}^{5}\text{He}^{\ast }+p,  \label{THe34} \\
&\rightarrow &\text{ }_{3}^{5}\text{Li}+n,  \label{THe35} \\
&\rightarrow &\text{ }_{3}^{5}\text{Li}^{\ast }+n,  \label{THe36} \\
&\rightarrow &\text{ }_{3}^{6}\text{Li}+\gamma \text{ \ \ \ \ \ \ \ \ \ \ \ }%
(Q=15.8\text{ MeV).}  \label{THe37}
\end{eqnarray}%
The amount of energy needed to fuse nuclei is proportional to the number of protons involved in the reaction. Because D$-$ $_{2}^{3}$He fusion involves 3 protons as opposed to 2 with D$-$T or D--D fusion, the amount of heat required for good fusion parameters is about 90 keV. This is about 10 times greater than the amount needed for D$-$T fusion. The measurements of the $ _{2}^{3}$He+$_{2}^{3}$He and T$+$ $_{2}^{3}$He fusion reactions in high-energy-density plasmas environment were reported recently \cite{zylstra2017proton}. It is worth mentioning that the $_{2}^{3}$He$+$ $_{2}^{3}$He and T$+$T reactions are mirror reactions, expected to be governed by similar nuclear physics after Coulomb corrections. In experiments \cite{casey2012measurements, sayre2013measurement} 
using high energy density plasmas environment were measured the neutron spectrum at very low center of mass energy 16--23 keV. 
The radiative capture reaction $_{2}^{3}$He($_{2}^{3}$He,$\gamma$)$^{6}$Be(\ref{He3He34}) has a branching ratio  $ \sim 4\times 10^{-5}$ and is thus negligible \cite{harrison1967radiative}. Among the primary reactions in D--$_{2}^{3}$He plasma our particular interest is addressed to the process (\ref{DD}), where 2.45 MeV neutrons are produced, and reactions (\ref{DD3He1}) and (\ref{DD1}), which are the
sources of tritium production. In hot D$-$ $_{2}^{3}$He plasma the D--T and T--T fusion admits reactions (\ref{DT})--(\ref{TT2}), where fast 2--14 MeV
useless and undesirable neutrons are emitted. As a result of T--$_{2}^{3}$He fusion the neutrons are emitted in reactions (\ref{THe32}), (\ref{THe35})
and (\ref{THe36}). Undesired neutron contamination is among the main risk factors for damaging of spacecraft materials due to a neutron radiation occurs as a result of the interaction of energetic neutrons with a lattice atom in the material and a crew during a mission. Thus the problem of the contamination due to the neutron emission in these processes exists if the
produced tritium is not removed to prevent its fusion with deuterium, helium-3 and itself. The tritium removal methods have been proposed in Refs.  \cite{khvesyuk1995ash, sawan2002impact}. Also the production of neutrons is reduced by rapidly exhausting tritium and by altering the fuel ratio to have three times the $_{2}^{3}$He as D, favouring the $_{2}^{3}$He reactions \cite{razin2014direct, cohen2015reducing} resulting in a great mass save.

The $^3$He fuel consumption is very complex to calculate and it depends on multiple factors. We base our estimate for fuel consumption on values calculated for other missions based on the DFD \cite{thomas2017fusion, paluszek2014direct}. The fuel consumption is calculated by dividing the total fuel consumption by the days of mission and the fusion power considered for the previous studies. In this way it has been found a fuel consumption per day per MW of power. This means that, for a 2.5 years mission, under the hypothesis of a 2-MW class DFD engine, the mass of $^3$He required would be about $0.27$ kg. This fuel mass value, on a spacecraft of multiple tonnes, can be neglected for all trajectories calculations. Although the amount of $^3$He on the surface of Earth is limited, as discussed in \cite{turner2003m}, this value is well below the maximum availability in human hands \cite{kennedy2018interstellar}. Moreover, there is a large source of extractable $^3$He ($10^9$ kg) on the lunar surface that have been deposited by solar wind for more than $4 \cdot 10^9$ years \cite{fa2010global,wittenberg1986lunar,fa2007quantitative}.

\subsection{Thrust model}
Researchers at PPPL performed simulations using UEDGE software \cite{mcgreivy2016}, a 2D multi-species fluid code, in order to model the cylindrically symmetric FRC open magnetic field plasma region of the DFD and find a steady-state self-consistent solution of continuity equations, momentum equations, and energy equations for each chemical species. This research allows to study the plasma parameters (temperature, density, velocity) and power flow within the SOL, each as a function of heating power and gas input \cite{mcgreivy2016}. The heated plasma will expand in a magnetic nozzle, converting its thermal energy into kinetic energy, thus providing thrust to the system. It works as a physical nozzle, with the difference that the fluid does not directly hit any physical wall. Researchers at Princeton Satellite Systems (PSS) considered this data to produce a functional model of the thrust and specific impulse of the engine as a function of input power to the SOL and propellant flow rate. It is essential to underline that the input power into the SOL is only $40$-$50 \;\%$ of the total fusion power. If only fusion products were ejected directly from the engine, they would have a velocity of $25,000$ km/s producing negligible thrust \cite{thomas2017fusion}, but interacting with cool ionized gas in the SOL, the energy is transferred from the hot products to the electrons and then transferred to the ions as they traverse the magnetic nozzle. The resulting exit velocity is about $10^5$ m/s, generating a thrust of about $2.5$ to $5$ N per MW of fusion power \cite{thomas2017fusion,cohen2017direct}, and from $5$ to $10$ N of thrust per MW of thrust power, with a specific impulse of about $10,000$ s. The estimated range for the specific power is between 0.75 kW/kg to 1.25 kW/kg. In Table \ref{DFDrange} the main characteristics for low and high power configurations of the engine are given. The DFD can be fully scaled in configuration and reach the power required. Let us assume the minimum estimated value of 0.75 kW/kg of specific power, which is a conservative option for all the calculations.
\begin{table}[H]
\caption{Direct Fusion Drive performance. The characteristics for low and high power configuration are shown \cite{cohen2017direct}.}
\centering
\begin{tabular}{r c c}
\hline \hline
     & Low power &  High power     \\
\hline
Fusion Power, [MW]         & 1  &      10       \\
Specific Impulse, [s]          & 8500 - 8000   &  12000 - 9900   \\
Thrust, [N]  &  4 \hspace{2mm} - \hspace{2mm} 5     &    35 \hspace{1mm} - \hspace{1mm} 55    \\
Thrust Power, [MW]          & 0.46    &      5.6     \\
Specific Power, [kW/kg]            &   0.75    &     1.25    \\
\hline \hline
\end{tabular}
\label{DFDrange}
\end{table}
Finally, due to the engine compactness, multiple modular DFDs can be combined into a cluster of many engines, resulting in a total thrust that is the sum of the single thrusts without affecting the $I_{sp}$.
\section{Earth - Titan mission} \label{mission}
Trajectory design for low-thrust propulsion systems represents a complex problem. As well known, spacecraft motion is governed by a sensitive system of non-linear differential equations and the inclusion of low-thrust forces into this system adds complexity to the trajectory determination problem. Therefore, it is convenient to consider the orbit as the evolution of an ellipse for each temporal instant defined by the instantaneous position and velocity vectors. Low-thrust effect causes the six formerly constant parameters to slowly vary from the Keplerian solution and it is possible to use the Gauss planetary equations to describe these rates of change \cite{vinti1973gaussian}. It is impossible to analytically evaluate the solution in the general case and the problem requires numerical methods to be solved. Several approaches are taken into account to further analyse potential trajectories to Saturn-Titan system using the Systems Tool Kit software from Analytical Graphics, Inc. The purpose of the research is to study the feasibility of such kind of mission with a $2$-MW class engine and to demonstrate the advantages related to this new propulsion concept. Let us consider that the thrust $T$ and specific impulse $I_{sp}$ are constant. The minimum estimated specific power of $0.75$ kW/kg is selected and the power to SOL is supposed of about $1$ MW, resulting in $8$ N of $T$ and about $9,600$ s of $I_{sp}$.

As a first step impulsive maneuvers are considered, computing the solution for the Lambert problem, determining the impulse that produces the orbit connecting a departure state (initial position and velocity) with a subsequent arrival point related to the target planet (final position). Lambert's law is used for an array of start dates and time of flights (between 2 and 5 years), and a minimum $ \Delta V$ is found. For instance, a 3-year trajectory with a start date in the next 30 years was found to require an impulsive $ \Delta V$ between $37$ and $45$ km/s, , depending on the date of departure. The planetary positions are obtained for the date given using JPL's HORIZONS system which can be used to generate ephemeris for solar-system bodies \cite{JPLHorizons}. It is essential to minimize the propellant mass required, because the unburnt propellant mass must be accelerated along the trajectory with the spacecraft itself. Therefore, an initial rough mass estimation is calculated through an iterative process on MATLAB using the Tsiolkovsky rocket equation \cite{dvornychenko1990generalized, kosmodemyansky2000konstantin}. As well known, this equation shows that for a rocket with a given empty mass and a given amount of fuel, the total change in velocity is proportional to the specific impulse. It is also important to consider that the increase of the mission duration, thrust efficiency, exhaust velocity, or specific power reduce the thrust duration and mass required for the mission. Increasing the payload has the reverse effect. A $25 \%$ burn duration of the 3-year time of flight is 280 days. A $ \Delta V$ of $40$ km/s can be achieved with a thrust of 8 N and a specific impulse of 9,600 s, with a total initial mass of less than $7000$ kg. These first rough estimates result in good initial guesses for the finite maneuvers analysis, which is essential to obtain accurate results for this kind of low-thrust engine. 
\section{Thrust-coast-thrust profile mission} \label{tctmission}
\begin{figure}[H]
  \centering
  \includegraphics[width=0.5\linewidth]{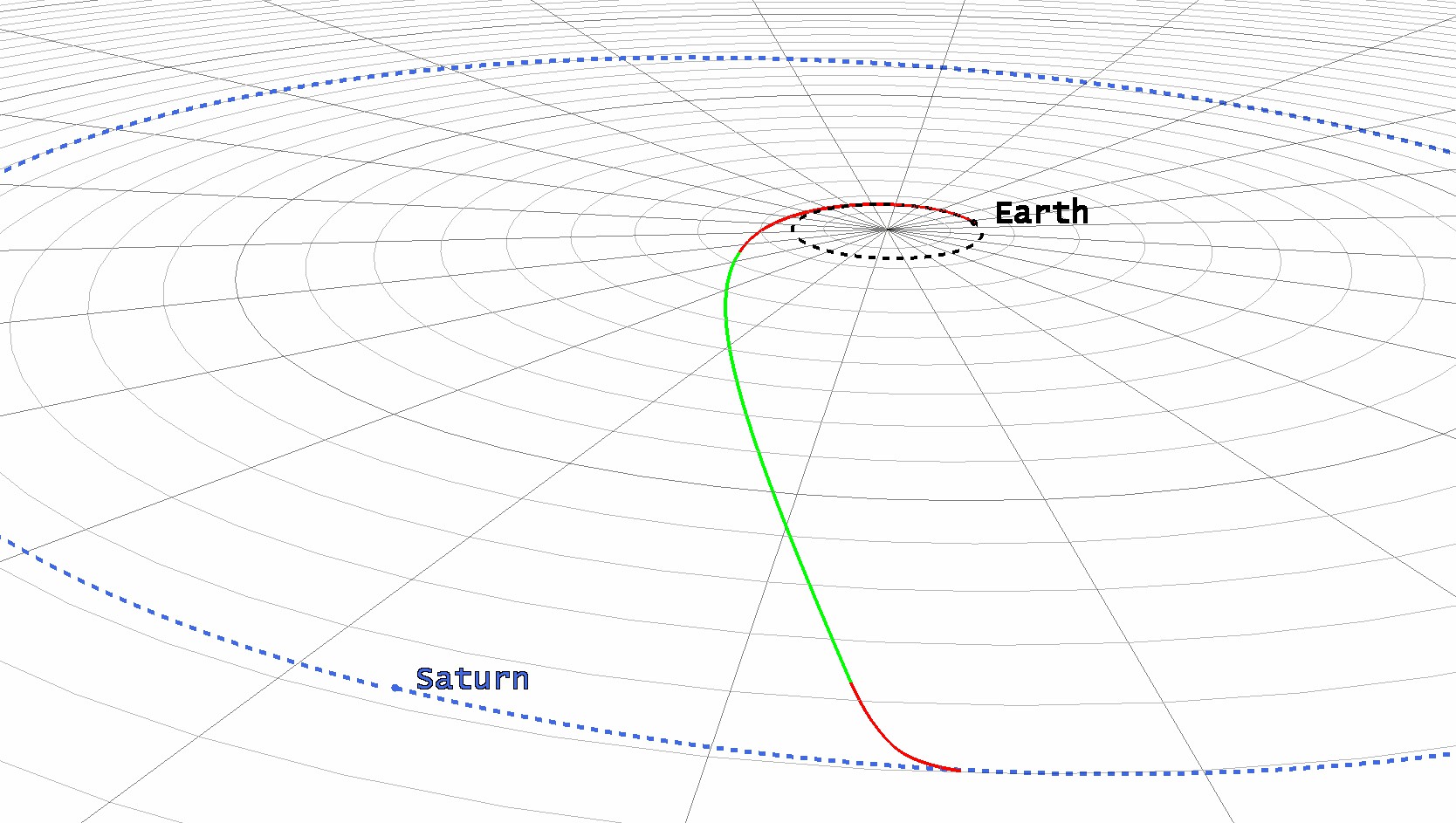}
  \caption{Thrust-coast-thrust profile for the Titan mission. It is possible to observe three segments of the trajectory, the red solid curves suggest that the spacecraft thrust is active and the green line represents the coasting phase without active thrust.}
  \label{fig:tct}
\end{figure}

The objective of the mission is to reach Saturn near the descending node referred to the ecliptic plane along its orbit around the Sun, in order to solve a nearly 2-D problem, with huge advantages at numerical and computational level. Once the spacecraft is orbiting around Saturn, the Titan orbit insertion maneuver concludes the mission. The T-C-T profile mission is divided into four different phases: Earth departure, interplanetary trajectory, Saturn orbit insertion and Titan orbit insertion. The mission start time has been estimated considering both the Earth and Saturn orbits and taking into account the time constraint represented by the rendezvous with the planet target. The inputs for this rendezvous problem are the DFD engine parameters, initial and final radii $r_1$ and $r_2$, payload and spacecraft mass (initial guesses). Travel time and mission start time estimations are obtained after several iterations, starting from initial guess related to the impulsive approach and considering some crucial constraints, for instance, related to the total mass of the spacecraft. Let us follow Fig. \ref{fig:tct}, where the first three phases of the mission are shown.
\noindent The first red solid curve starts from the Earth initial position and contains both the escape maneuver from Earth and the burn which puts the spacecraft along the heliocentric hyperbolic trajectory to Saturn. The second red solid curve shows the Saturn orbit insertion. While the green curve indicates the interplanetary trajectory without thrust. In the following subsections each phase is analyzed.
\subsection{Earth departure} \label{ter}
A logic solution for the Earth departure phase could be to insert the spacecraft directly into a heliocentric orbit. Our simulations show that the Earth departure from a low Earth orbit (LEO) uses reasonable propellant mass and takes between 25 and 71 days depending on the DFD engine parameters and initial mass considered. The results obtained are very close to those published in Ref. \cite{thomas2017fusion}. This solution allows to use almost any launch vehicle, dramatically reducing launch and overall mission costs. The initial considered orbit is a circular orbit with an altitude of about 386 km and inclination of about 24 degree which allows the spacecraft to escape from Earth gravitational influence along the Ecliptic plane.  A simulation for the spiral trajectory is presented to evaluate propellant mass consumption and maneuver time. As a first step, an Earth point mass model is considered and we neglect all orbital perturbations. This is a representative model because the influence of all orbital perturbations leads only to a small difference in terms of propellant mass and maneuver duration. The results of calculations related to the Earth escape maneuver are listed in Table \ref{fig:OPTIMEscapeMan}. We approximated the optimal thrust steering law \cite{boltz1992orbital} with a model where the thrust is mainly aligned with the velocity vector. However, the thrust vector has a small positive radial component too.
\begin{table}[H]
\caption{Earth departure spiral analysis. The final results are obtained considering the Earth point mass model and a geocentric reference system.}
\centering
\begin{tabular}{r c}
\hline \hline
 & Departure phase \\
\hline
Mission start time & 2 Nov 2046   \\
Maneuver duration & 76.2 days         		\\
Payload mass & 1800 kg             	  \\
Initial mass & 7250 kg               \\
Propellant consumption   & 559.5 kg               \\
Van Allen belt time      & 17 days\\
$\Delta V$ & 7.562 km/s             \\
Initial velocity & 7.676 km/s             \\
Final velocity & 1.518 km/s             \\
\hline \hline
\end{tabular}
\label{fig:OPTIMEscapeMan}
\end{table}
 It is essential to emphasize that the spacecraft should escape from Earth with a velocity vector parallel to the Earth's orbital velocity (with respect to the Sun) before entering the interplanetary space. This condition is crucial to take full advantage of the Earth's orbital speed ($V_{CE_1} \approx 30$ km/s) and it is achieved acting on the waiting time in LEO. Therefore, as a consequence of a long iterative process which depends by several variables involved in all mission phases, the maneuver start time is obtained with a given waiting time in LEO.
\begin{figure}[H]
  \centering
  \includegraphics[width=0.6\linewidth]{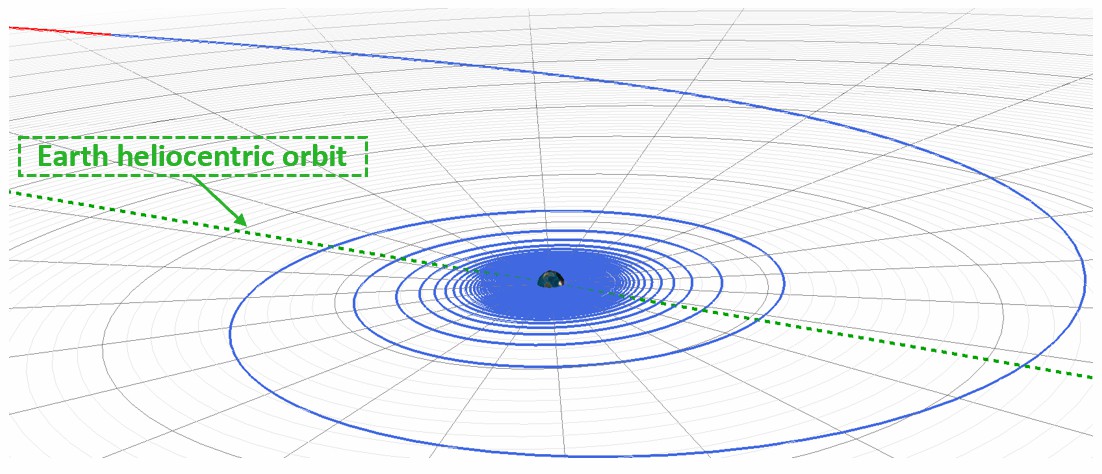}
  \caption{Spiral trajectory for the escape maneuver. The escape trajectory (blue) is shown using the Earth centred reference system. The red solid line starts when the spacecraft is outside the Earth gravitational sphere of influence, moving into interplanetary space.}
  \label{fig:Escape}
\end{figure}
Due to the electromagnetic nature of the engine and because of safety reasons for the spacecraft itself, it is preferable to spend less time possible inside the inner Van Allen radiation belt. This time is about 17 days and it is calculated as the time spent between $1,000$ and $6,000$ km altitude. In order to simulate the entire Earth departure phase the analysis stops when the spacecraft reaches the external surface of the Earth's gravitational sphere of influence (radius $r_{\oplus_{ \infty}}\approx 10^6$ km). The escape maneuver ends when the eccentricity reaches the unit value after $71$ days of burn with a propellant consumption of about $520$ kg.
\subsection{Interplanetary trajectory}
Once the Earth departure phase is completed, the spacecraft is in an elliptic orbit around the Sun and the interplanetary trajectory mission phase begins. Many days of acceleration are needed at this point to obtain an heliocentric hyperbolic orbit, where the thrust vector is aligned with the velocity of the spacecraft. This is essentially the continuation of the previous maneuver, since the engine still generates thrust while it is exiting from the Earth gravitational sphere. It comes from an iterative process where the launch date, maneuver duration and thrust direction components are the main independent variables. After the acceleration phase, it is necessary to include a proper coasting segment, which duration has a strong impact on the following maneuvers. The main goal of this phase is to reach a spatial region on the ecliptic plane, pointing at the descending node of Saturn's heliocentric orbit. It is necessary to include a 1.6 year long coasting segment before the deceleration maneuver when the spacecraft is approaching the target planet, ensuring that the spacecraft and Saturn velocities are comparable. The parameters related to this phase are shown in Table \ref{DSM1}.
\begin{table}[H]
\caption{Characteristics of the interplanetary phase. The results related to the finite maneuver after the Earth escape are listed in the upper part of the table. The second half shows the coasting phase, which starts when the spacecraft shuts down the thrust generation. The velocities are shown in an heliocentric reference system.}
\centering
\begin{tabular}{r c}
\hline \hline
 & Acceleration phase \\
\hline
Maneuver duration & 130.1 days         		\\
Propellant consumption     & 955.18 kg               \\
$\Delta V$ & 14.503 km/s             \\
Initial velocity & 31.786 km/s             \\
Final velocity & 34.560 km/s             \\
\hline \hline
 & Coasting phase \\
\hline
Coasting duration & 1.67 years    		\\
Initial velocity  & 34.560 km/s             \\
Final velocity & 19.659 km/s             \\
\hline \hline
\end{tabular}
\label{DSM1}
\end{table}

\subsection{Saturn orbit insertion} \label{insert}
\begin{figure}[H]
  \centering
  \includegraphics[width=0.35\linewidth]{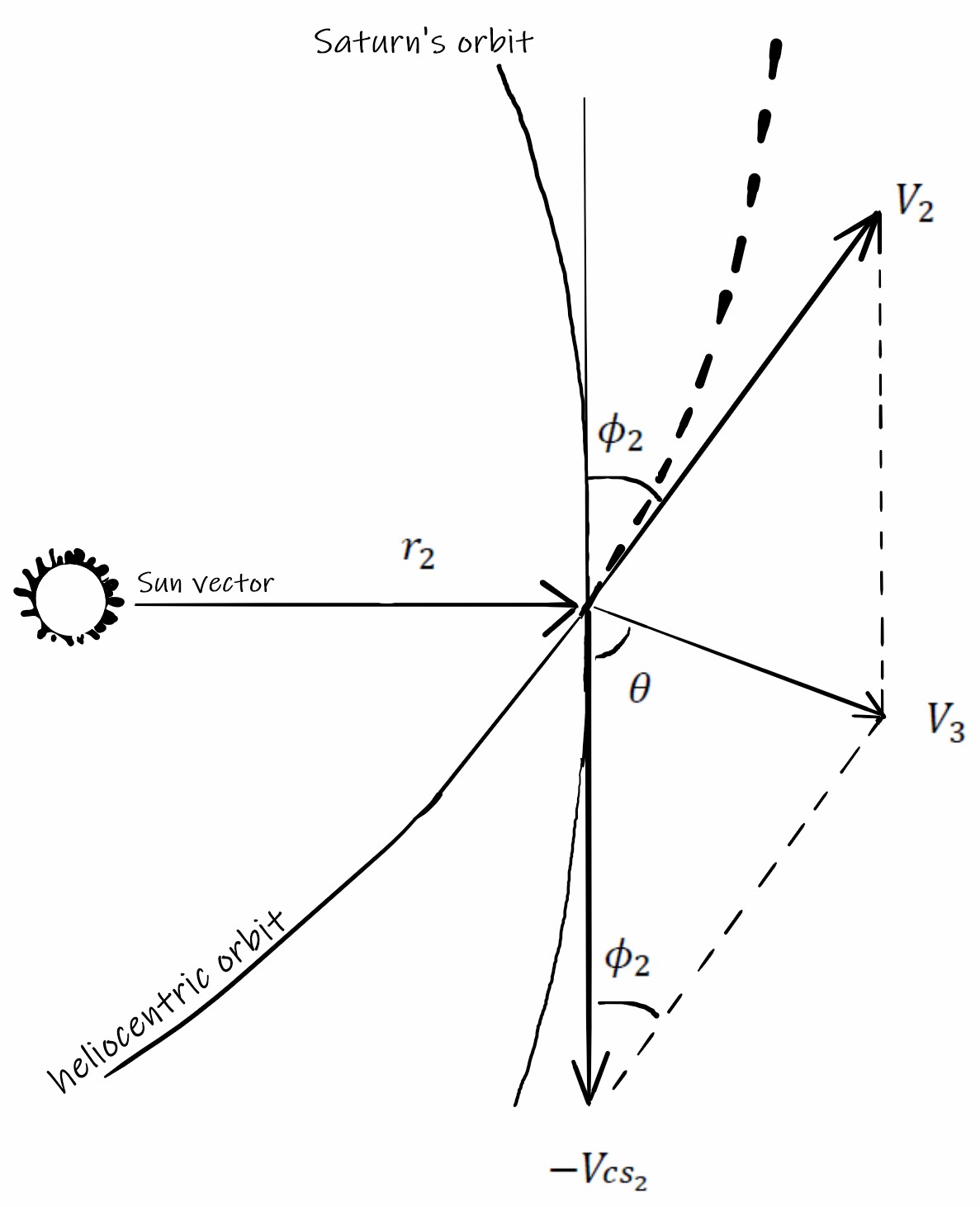}
  \caption{Fundamentals parameters for rendezvous mission. $\phi_2$ is the angle between the spacecraft heliocentric velocity vector $V_2$ and the orbital velocity of the target planet at arrival $V_{CS_2}$. $V_3$ is the velocity of the spacecraft relative to the planet target.}
  \label{fig:angolo}
\end{figure}

The Saturn orbit insertion (SOI) maneuver puts the spacecraft into Saturn's orbit. 
The main purpose of the SOI is to obtain a proper velocity vector which allows the spacecraft to orbit around Saturn at a radius comparable to that of Titan. Let us adopt a Saturn centred reference system. It is required that after the SOI maneuver the spacecraft orbit switches from a hyperbolic trajectory to an elliptical orbit around the target planet. At arrival, however, the heliocentric transfer orbit usually crosses the target planet's orbit at some angle, $\phi_2$ as shown in Fig. \ref{fig:angolo}. The heliocentric spacecraft velocity, $V_2$, and the orbital speed of the target planet, $V_{CS_2}$, should be comparable in magnitude with a relatively small angle $\phi_2$ between them. Otherwise, only a trajectory variation occurs (fly by or gravity assist) with a net accelerative effect which depends on the angle $\theta$ and the velocity of the spacecraft relative to the target planet $V_3$ \cite{bate2019fundamentals}. As a first step, a simple impulsive problem allows to evaluate the hyperbolic excess velocity $V_{3_ \infty}$ on the approach hyperbola to Saturn and other physical parameters. This important estimations are necessary to have accurate initial guesses of the velocities magnitude and periapsis radius of the approach trajectory, for the finite numerical analysis. It is essential to consider the Titan orbital parameters in order to initially define the periapsis radius $r_p$ of the approach trajectory. As the distance of closest approach to Saturn we consider the mean distance from the center of Saturn to the Titan, assuming a comparable value $r_p \approx 3 \cdot 10^6$ km. An iterative process is used to estimate the velocity $V_2$ when entering in the Saturn influence, which is clearly bound to the spacecraft velocity $V_3$ relative to Saturn. A good initial guess for the arrival velocity is slightly higher to the descending node Saturn's orbital speed ($V_{CS_2} = 9.17$ km/s).
\noindent The necessary change in velocity is not achievable only decelerating the spacecraft, but also varying the direction of $V_2$ acting on the radial component of the thrust. Since $V_3$ is the result of the cross product ($V_{CS_2}$ and $V_2$), if the $\phi_2$ angle decreases, the resulting hyperbolic excess velocity rotates and decreases in magnitude, as noticeable in Fig. \ref{fig:angolo}. Finally, analysing the problem taking into account finite maneuvers, both deceleration and direction changes are studied. In order to avoid an excessive deceleration, which would extend the mission time, a trade-off between time requirements and propellant mass consumption allows to find a solution. The SOI maneuver is divided in two segments. Near the end of the first segment the spacecraft is entering the planet gravitational sphere of influence and starts to be increasingly attracted.
\begin{table}[H]
\caption{Approach to Saturn. Results related to the entire Saturn orbit insertion maneuver in a heliocentric system.}
\centering
\begin{tabular}{r c}
\hline \hline
& SOI maneuver \\
\hline
Maneuver duration & 142.2 days         		\\
Propellant consumption  & 1044 kg               \\
$\Delta V$ & 18.901 km/s             \\
Initial velocity  & 19.659 km/s             \\
Final velocity  & 13.780 km/s             \\
\hline \hline
\end{tabular}
\label{DSM2}
\end{table}

\begin{figure*} [ht]
\centering
{\includegraphics[width=8cm]{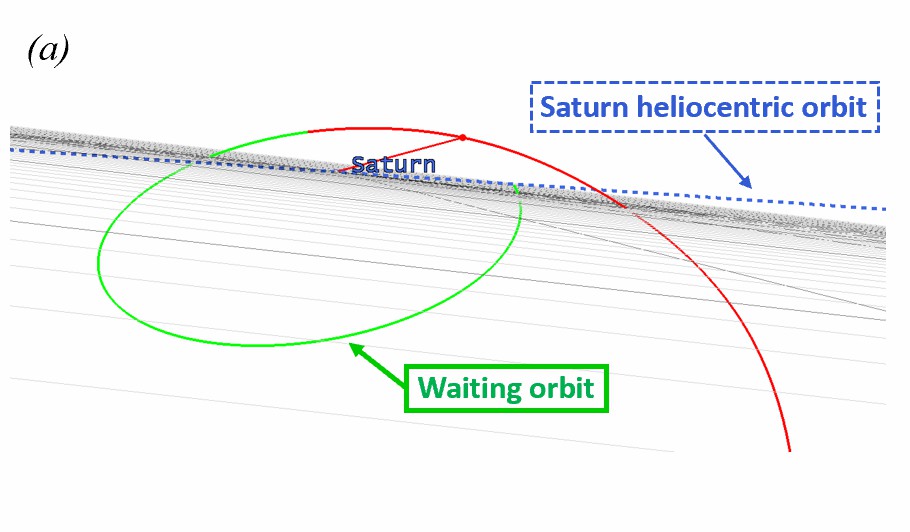}}
\hspace{2mm}
{\includegraphics[width=8cm]{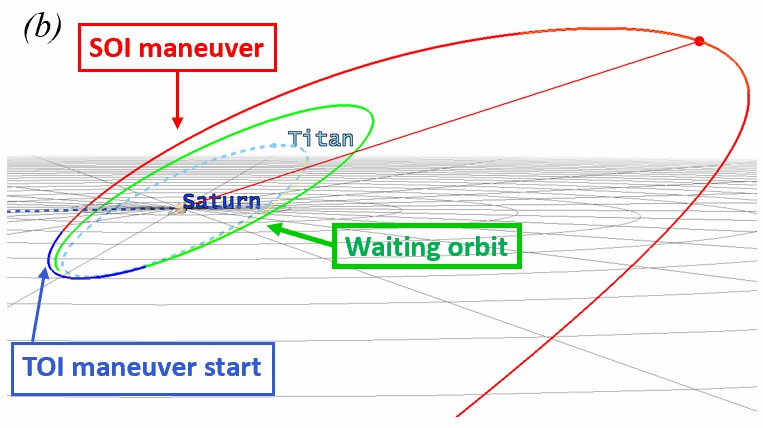}}
\caption{Saturn orbit insertion maneuvers. Figure (\textit{a}): One of the solution for the SOI maneuver. The red solid curve represents the finite maneuver which let the spacecraft to orbit around Saturn on the same orbital plane of Titan. The green solid line is the waiting orbit, leading up to the the TOI maneuver. Figure (\textit{b}) Alternative solution, the red solid curve represents the finite maneuver which let the spacecraft to orbit around Saturn on the same orbital plane of Titan. The blue solid line marks the start of the TOI maneuver.}
\label{fig:insertion_1}
\end{figure*}

It is worth to underline that the spacecraft approaches Saturn shortly after the descending node, achieving similar orbital parameters with respect to Titan. This choice allows to reduce the propellant consumption, taking advantage of Saturn's vertical motion relative to the ecliptic plane. Notice that the SOI maneuver has only a small normal component which allows to achieve the same inclination and right ascension of the ascending node of Titan orbit, as shown in Fig. \ref{fig:insertion_1}. Finally, when the spacecraft orbits around Saturn on the same orbital plane as Titan, the problem is once again bi-dimensional.
Once the spacecraft is captured by Saturn, a possible solution is to put the spacecraft into a proper orbit to wait the optimal time to start the next DSM which leads to a Titan centred orbit. During the SOI maneuver, precious scientific data related to Saturn's magnetosphere, rings and other interesting scientific objectives \cite{harland2002mission} can be collected. For this reason, we orbit the main body in an elliptical waiting orbit with a semi-major axis of about $1.8 \cdot 10^6$ km, for about 170 days before starting the last mission maneuvers.
\subsection{Titan orbit insertion}
\begin{figure}[H]
\centering
\includegraphics[width=0.55\linewidth]{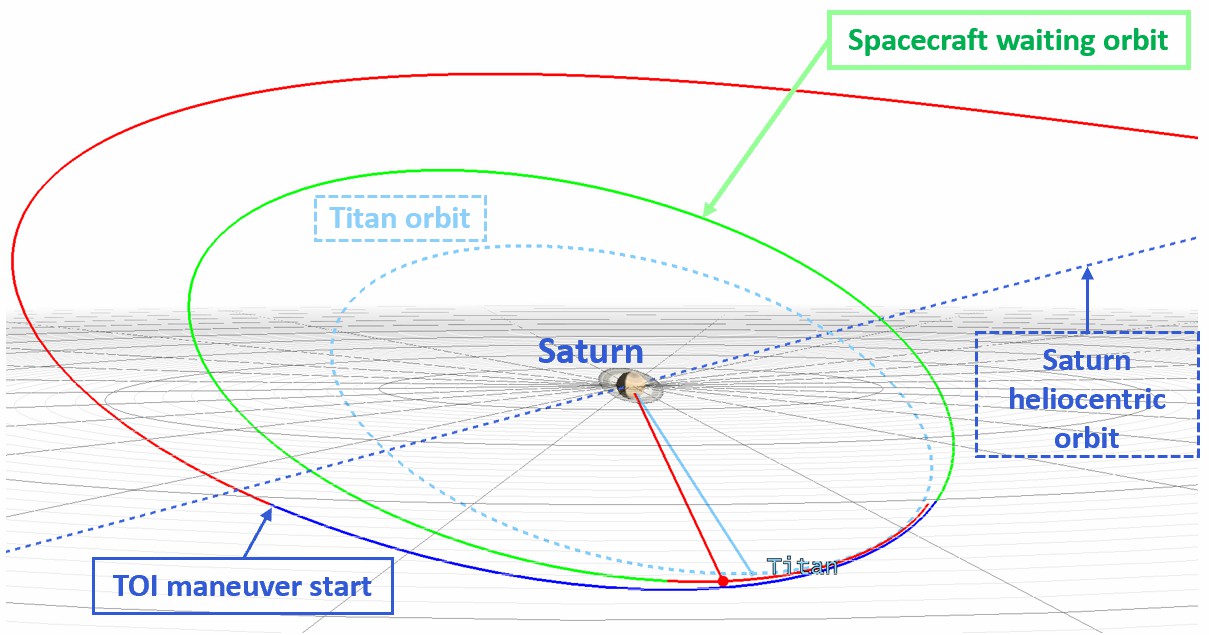}
\caption{Titan orbit insertion phase. Central body: Saturn. The blue line marks the start of the required TOI maneuver, subsequent the SOI maneuver (red). It is possible to observe the portion of the waiting orbit (green) travelled by the spacecraft before the phasing maneuver starts (red).}
\label{fig:phasing_1}
\end{figure}

\begin{figure}[H]
  \centering
  \includegraphics[width=0.6\linewidth]{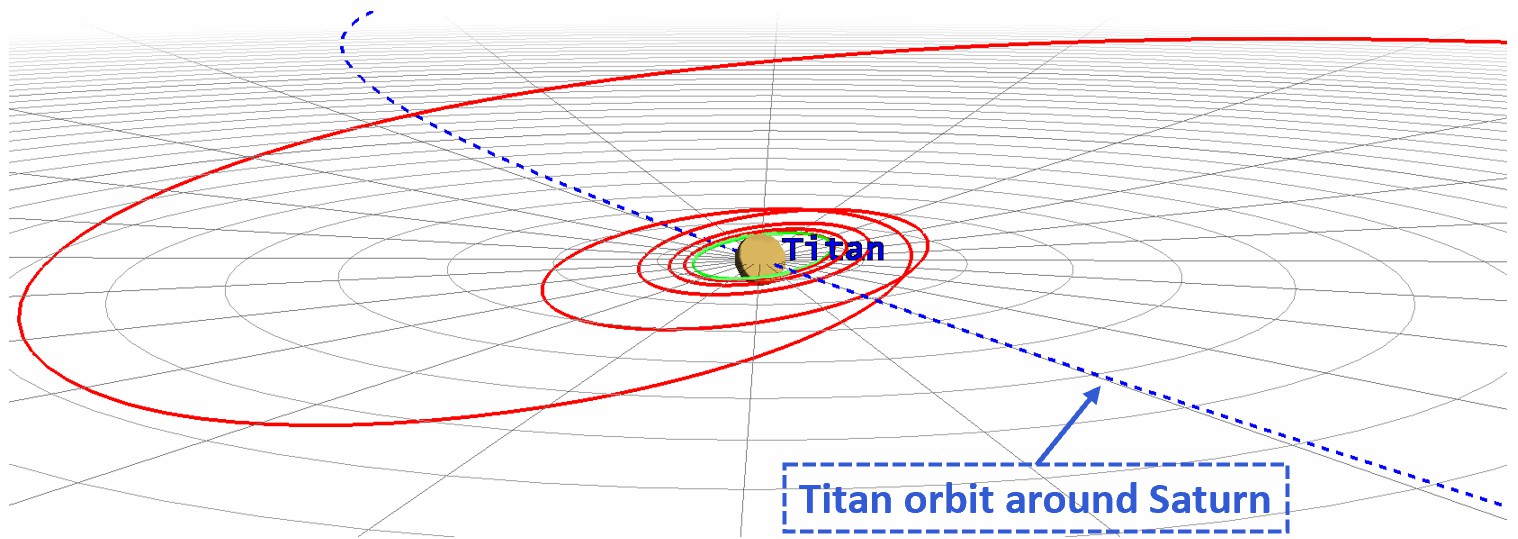}
  \caption{Titan orbit insertion from a Titan centred inertial reference system. After the phasing maneuver, the red solid curve represents the orbit insertion maneuver which results in the final orbit around Titan (green solid curve).}  
  \label{fig:insertionTitan_3}
\end{figure}

Different solutions are considered also for the Titan orbit insertion (TOI) phase. In order to properly deal with the mutual gravitational interaction between Saturn, Titan and the spacecraft it is necessary to address the three-body problem. This problem consists of determining the perturbations in the motion of one of the bodies around the central body, produced by the attraction of the third. The Moon's orbit around the Earth, as disturbed by the Sun is an example. In order to numerically solve the problem and obtain accurate results we use a perturbation model that takes into account the gravitational influence of both Saturn and Titan. The purpose of this phase is not only to achieve orbital parameters very similar to those of Titan, but also to perform a rendezvous. Therefore, a strong time constraint has to be considered to solve the problem. Let us adopt the following maneuver strategy: i) a  first finite burn where the thrust vector is directed along the anti-velocity direction; ii) a coasting segment; iii) a phasing maneuver
where thrust components along anti-velocity
and co-normal direction are considered; iv) the final Titan orbit insertion maneuver. The different segments of the TOI maneuver can be seen in Fig.\ref{fig:phasing_1}. The phasing angle, which is the angle between the spacecraft and Titan position vectors, with respect to Saturn, decreases during this entire maneuver because of the different orbits travelled by the chaser (spacecraft) and the target (Titan) resulting in different velocities, as shown in Fig. \ref{fig:phasing_1}.
\begin{table}[H]
\caption{Titan orbit insertion phase. Results related to the required insertion maneuver which let the spacecraft (chaser) reaches Titan (target), orbiting around Saturn. The results are related to a Saturn centred reference system.}
\centering
\begin{tabular}{r c}
\hline \hline
 & TOI maneuver \\
 \hline
Maneuver duration & 33.5 days         		\\
Propellant consumption  & 77 kg               \\
$\Delta V$  &  1.569 km/s             \\
Initial velocity & 4.163 km/s             \\
\hline \hline
\end{tabular}
\label{tab:phasingTOT}
\end{table}
\noindent Once the spacecraft reaches Titan with the proper relative velocity, the final orbit insertion maneuver starts, achieving a Titan centred circular orbit 4000 km away from the surface, as shown in Fig. \ref{fig:insertionTitan_3}. This altitude provides sufficient orbital stability, requiring less station keeping maneuvers to fight Saturn's perturbations, which can be useful to freely vary the spacecraft orbit around Titan. Moreover, in order to meet the scientific instruments requirements, it is possible to achieve closer distance to the surface by varying thrust direction and duration with a negligible amount of propellant.
In summary, the T-C-T mission profile is based on the assumption that the DFD will be capable of turning off and on the thrust generation. This is an important hypothesis which requires that the engine will not produce thrust during the coasting phase, which is in theory possible but not yet certain. More specifically, because of the robotic nature of the mission, it could be possible to think to turn off the engine in order to save both deuterium propellant and precious fuel ($^3$He). The total fuel consumption for the entire mission is $\approx 0.112$ kg. Another possible solution, which could be more feasible, is based on the DFD ability to turn off and on the thrust generation without shutting down the engine, still generating the electrical power from the nuclear fusion reactions. In this case the reactor still provides energy for the entire mission, and the $^3$He consumption would be around $0.282$ kg.
\section{Continuous thrust profile mission} \label{ct}
The natural alternative solution to the T-C-T mission profile is represented by the continuous thrust (CT) profile mission where the engine is always on, generating a constant thrust during the previous four different mission phases.
\begin{figure}[H]
  \centering
  \includegraphics[width=0.5\linewidth]{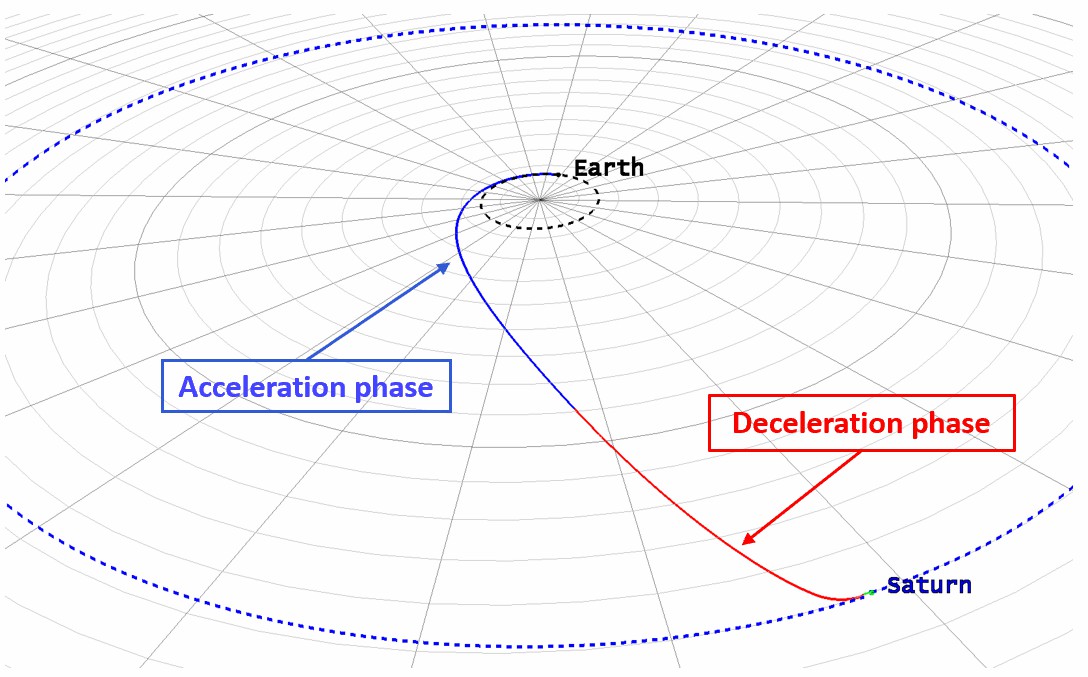}
  \caption{Planar trajectory for the continuous thrust profile mission. At the end of the blue curve there is the change in direction of the thrust (switch time). The trajectory follows Earth’s orbit for some time before a nearly straight trajectory to Saturn.}
  \label{fig:profiloTcontinua}
\end{figure}
\noindent An iterative process is necessary to define the spacecraft initial mass, in order to define the propellant mass for the entire mission, which is directly related to the total duration of the trip. Starting from the same orbital initial conditions of Section \ref{ter}, calculations are performed with the additional constraint on the thrust. The results from the first phase analysis are listed in Table \ref{datiterra}.
\begin{table}[H]
\caption{Earth departure spiral analysis. The results are obtained considering an Earth point mass model and a geocentric reference system}
\centering
\begin{tabular}{r c}
\hline \hline
 & Departure phase \\
 \hline
Mission start time & 25 Sep 2047   \\ 
Maneuver duration & 93 days         		\\
Payload mass & 1000 kg             	  \\
Initial mass & 9015 kg               \\
Propellant consumption   & 686.99 kg               \\
$\Delta V$ & 7.475 km/s             \\
Initial velocity  & 7.674 km/s             \\
Final velocity  & 1.376 km/s             \\
\hline \hline
\end{tabular}
\label{datiterra}
\end{table}
The main goal of this scenario calculation is to find a proper switch time for the thrust direction inversion in order to reach Saturn with an acceptable velocity, allowing the spacecraft to orbit the planet. The same philosophy leads to the Saturn and Titan orbit insertion maneuvers, which conclude the analysis and the related results are listed in Table \ref{datisat}.
\begin{table}[H]
\caption{Interplanetary phase. The results are related to the acceleration and deceleration phases for the heliocentric orbit.}
\centering
\begin{tabular}{r c}
\hline \hline
 & Acceleration phase \\
\hline
Initial mass & 8313 kg              \\
Burn duration & 383.05 days \\
Propellant consumption     & 2812 kg  \\
$\Delta V$  & 14.503 km/s   \\
Initial velocity & 31.649 km/s             \\
Velocity at switch time & 47.814 km/s             \\
\hline \hline
 & Deceleration phase \\
\hline
Burn duration  &  238 days   	\\
Propellant consumption     & 1749 \\ 
$\Delta V$ & 36.035 km/s             \\
Final velocity  & 12.509 km/s             \\
\hline \hline
\end{tabular}
\label{datisat}
\end{table}
\section{Scenarios comparison} \label{confronto}
Let us consider the two mission scenarios analysed, the T-C-T and CT, and compare them to one of the most successful scientific mission, Cassini-Huygens, which studied the Saturn system for more than 10 years. A summary for the entire scenarios studied are given in order to make a comparison of mission durations, payload masses and propellant consumptions. In Table \ref{comparison1vera} along with results of our calculations are also presented data for the Cassini-Huygens mission taken from Refs. \cite{mitchell2006cassini, NASA, ESA, johnson2005power}. 
\begin{table*} [ht]
\caption{Comparison between the thrust-cost-thrust profile, the continuous thrust profile and Cassini-Huygens missions. The CT profile mission results into an even shorter time travel - less than two years - still with a heavier payload than previous missions.}
\centering
\begin{tabular}{l c c c }
\hline \hline
		& \textbf{T-C-T profile} & \textbf{CT profile} & \textbf{Cassini-Huygens mission} \\
\hline
Travel time (to Saturn), [days] &  958.50 
&  714.05  & 2422 \\ 
Initial spacecraft mass, [kg] & 7250 &  9015 & 5712 \\
Payload mass, [kg] & 1800  &  1000 & 617.4 \\
Propellant mass used, [kg]  & 2658 &  5347 & 2950 \\ 
Fuel ($^3$He) mass used, [kg]  & 0.282 & 0.201 & --\\
Maximum trip velocity (Sun), [km/s] & 34.560 & 47.814 & --\\
\hline \hline
\end{tabular}
\label{comparison1vera}
\end{table*}
It is worth noticing that in the CT case, the DFD is capable of really fast travels, rapidly reaching extremely high velocity. In the last mission phases, when a great amount of propellant has already been consumed and the spacecraft mass is decreased, the DFD can reach the required speed in a relatively short period, reducing of many years the time of flight with a payload decrease since the propellant consumption has doubled. The reduction in payload mass can be explained looking at the Earth departure phase: the higher the initial mass, the longer the time necessary to escape from Earth. Then, a payload of about 1000 kg has been obtained through an iterative process, in order to provide a relatively fast spiral Earth departure comparable with the first scenario solution. Collectively, such kind of maneuvers would be too demanding for any kind of present propulsion systems. The total trip duration is below 2 years for the CT profile mission, which is more than three times less the duration of the Cassini spacecraft travel to Saturn, which has been possible due to several gravity assists. It is important to emphasize that also the payload has increased significantly, delivering 1000 kg in the fastest solution or 1800 kg in the T-C-T profile mission. For comparison, the Cassini spacecraft had a total payload of about 617 kg, including the Huygens lander (349 kg) \cite{harland2002mission}. Another advantage related to the shortening of total mission duration is the reduction of precious fuel mass ($^3$He), compared with the scenario of the T-C-T profile mission. In the case of a robotic mission, such as Cassini mission, it could be possible to shut down the engine, saving precious $^3$He fuel. Otherwise, for a manned space mission this could not be reasonable due to the fact that the electrical power generation could be vital for the crew.

Finally, there are two possible solutions related to the operative phase of the mission. The results lead to two different feasible mission concepts. The high payload capability for both mission profiles allows to consider a parachute descent through Titan's dense atmosphere performed by a lander probe, containing a rover or even better a rotorcraft, carried on the main spacecraft (orbiter) \cite{tobie2006episodic,elachi2005cassini}. In this case, during the TOI maneuver, the orbiter will release the lander and keep orbiting around Titan or, by performing a proper maneuver, it can orbit again around Saturn. In the first case, the lander can be designed to collect scientific data for the entire mission, sending it to the powerful orbiter that is capable to receive and retransmit back data to Earth. In the second case, the lander will send data to the orbiter only during the atmospheric descent and a period after the landing, limited by the orbiter spacecraft trajectory and the lander power capability. Therefore a maneuver that changes the path of the orbiter requires a negligible $ \Delta V$ of tens m/s order of magnitude to perform the data relay during the descent.
\section{Conclusions} \label{conclusioni}
Realistic trajectories analysis to accomplish a rendezvous interplanetary mission 
are presented. Our analysis confirm that a Titan mission using a 2-MW class DFD engine is not only feasible, but the departure from LEO dramatically reduces launch and overall mission costs. The strong advantages related to this new propulsion technology result in a great reduction of travel time with respect to the previous performed missions \cite{acuna1980magnetic, smith1981encounter, smith1982new, harland2002mission} and a tremendous payload increase with a huge availability of on-board electrical power. The DFD would be a true game-changer for any robotic missions to asteroids, solar system planets and moons, and any other deep space mission become faster and cheaper. There are many missions that can be accomplished now with a small amount of $^3$He from terrestrial sources, and enormous reserves are presumably available on the Moon for future missions \cite{wittenberg1986lunar,fa2007quantitative}. Results of calculations obtained are extremely promising although the conservative assumptions on the DFD engine specific power. We estimate the mission phases duration and the propellant mass consumption for all the required maneuvers. In order to accomplish this goal, the proper thrust vector orientation and maneuver duration are numerically estimated. Most of the time the thrust vector is considered updated throughout the maneuver to maintain the required thrust direction. This choice forces the thrust vector to the desired direction at every instant during the burn, rotating with a specified coordinate system or tracking with the spacecraft's inertial velocity vector. We demonstrate that the total Earth-Titan mission duration is about 2.6 years for the T-C-T profile, and less than 2 years for CT profile, which is more than three times less the duration of the Cassini spacecraft travel to Saturn.

In future research works, it could be more appropriate to consider an \textit{inertial at ignition} condition during the insertion phases, where the thrust vector direction is defined at ignition and remains the same through the maneuver. This option does not require a continuous attitude change and it could make the maneuver more simple. However, this choice does not affect significantly the results. We performed a basic optimization process, which has proven to be useful to approximate the optimal thrust direction, minimizing the duration of burns, hence the propellant used. The payload increase, combined with the huge electrical power availability generated by the fusion reactor, leads to a tremendous growth of scientific data. In fact, for any robotic mission, the higher the payload, the more scientific instruments can be carried on-board and the more precious data can be collected. This is a key aspect, also thinking to the near future lunar and Mars missions \cite{genta2020achieving, smith2020artemis, ESAweb} where it would be essential to maximize the payload, without excessively extend the journey time.

\section*{Acknowledgements}
\noindent We thank S. A. Cohen and his research team at Princeton Plasma Physics Laboratory, who provided insight and expertise that greatly assisted the research.

\bibliographystyle{elsarticle-num}
\biboptions{sort&compress}
{\setstretch{0.1}\bibliography{MG}}

\begin{thebibliography}{10}
\expandafter\ifx\csname url\endcsname\relax
  \def\url#1{\texttt{#1}}\fi
\expandafter\ifx\csname urlprefix\endcsname\relax\def\urlprefix{URL }\fi
\expandafter\ifx\csname href\endcsname\relax
  \def\href#1#2{#2} \def\path#1{#1}\fi

\bibitem{cohen2017direct}
S.~A. Cohen, C.~Swanson, et~al., {Direct fusion drive for interstellar
  exploration}, J. Br. Interplanet. Soc. 72 (2019) 37--50.

\bibitem{godwin1999apollo}
R.~Godwin, Apollo 11: The NASA Mission Reports, Apogee Books, Burlington, 1999.

\bibitem{launius1997chronology}
R.~D. Launius, A chronology of mars exploration, in: NASA Technical Report,
  1997.

\bibitem{branigan1965mariner}
T.~L. Branigan, Mariner 4: Mission to mars, Phys. Teach. 3 (1965) 303--307.

\bibitem{chappell1995mars}
D.~T. Chappell, Mars subsurface radar mapper, J. Br. Interplanet. Soc. 48
  (1995) 395--404.

\bibitem{arbit1970investigation}
H.~Arbit, S.~Clapp, C.~Nagai, Investigation of the lithium-fluorine-hydrogen
  tripropellant system, J. Spacecr. Rockets 7 (1970) 1221--1227.

\bibitem{aubert2005cardiovascular}
A.~E. Aubert, F.~Beckers, B.~Verheyden, Cardiovascular function and basics of
  physiology in microgravity, Acta cardiol. 60 (2005) 129--151.

\bibitem{samarabandu1993analysis}
J.~Samarabandu, R.~Acharya, et~al., Analysis of bone x-rays using morphological
  fractals, IEEE trans med imaging 12 (1993) 466--470.

\bibitem{genta2020achieving}
G.~Genta, R.~Ya.~Kezerashvili, Achieving the required mobility in the solar
  system through direct fusion drive, Acta Astronaut. (2020) 303--309.

\bibitem{AimeGajeri2020}
P.~Aime, M.~Gajeri, R.~Ya.~Kezerashvili, Exploration of trans-neptunian objects
  using the direct fusion drive, Acta Astronaut. 178 (2021) 257–264.

\bibitem{romanelli2005assessment}
F.~Romanelli, C.~Bruno, G.~Regnoli, Assessment of open magnetic fusion for
  space propulsion, ESA-ESTEC Final Report 18853 (2005).

\bibitem{robbins1991historical}
W.~Robbins, An historical perspective of the nerva nuclear rocket engine
  technology program, in: Conference on Advanced SEI Technologies, (1991).

\bibitem{razin2014direct}
Y.~S. Razin, G.~Pajer, et~al., A direct fusion drive for rocket propulsion,
  Acta Astronaut. 105 (2014) 145--155.

\bibitem{cohen2011rf}
S.~Cohen, C.~Brunkhorst, et~al., Rf plasma heating in the pfrc-2 device:
  Motivation, goals and methods, in: AIP Conference Proceedings, Vol. 1406,
  2011, pp. 273--276.

\bibitem{thomas2017fusion}
S.~J. Thomas, M.~Paluszek, S.~Cohen, et~al., {Fusion-enabled Pluto Orbiter and
  Lander}, Princeton Satellite Systems, Inc. ((2019) Technical report).

\bibitem{cohen2000ion}
S.~A. Cohen, A.~H. Glasser, Ion heating in the field-reversed configuration by
  rotating magnetic fields near the ion-cyclotron resonance, Phys. Rev. Lett.
  85 (2000) 5114.

\bibitem{cohen2007stochastic}
S.~A. Cohen, A.~Landsman, A.~Glasser, Stochastic ion heating in a
  field-reversed configuration geometry by rotating magnetic fields, Phys.
  Plasmas 14 (2007) 072508.

\bibitem{kolb1959field}
A.~Kolb, C.~Dobbie, H.~Griem, Field mixing and associated neutron production in
  a plasma, Phys. Rev. Lett. 3 (1959) 5--8.

\bibitem{cohen2000maintaining}
S.~A. Cohen, R.~Milroy, Maintaining the closed magnetic-field-line topology of
  a field-reversed configuration with the addition of static transverse
  magnetic fields, Phys. Plasmas 7 (2000) 2539.

\bibitem{cohen2007formation}
S.~A. Cohen, B.~Berlinger, C.~Brunkhorst, et~al., Formation of collisionless
  high-$\beta$ plasmas by odd-parity rotating magnetic fields, Phys. Rev. Lett.
  98 (2007) 145002.

\bibitem{guo2005observations}
H.~Guo, A.~Hoffman, L.~Steinhauer, Observations of improved confinement in
  field reversed configurations sustained by antisymmetric rotating magnetic
  fields, Phys. Plasmas 12 (2005) 062507.

\bibitem{rosenbluth1979mhd}
M.~Rosenbluth, M.~Bussac, {MHD} stability of spheromak, Nucl. Fusion 19 (1979)
  489--498.

\bibitem{glasser2002ion}
A.~H. Glasser, S.~A. Cohen, {Ion and electron acceleration in the
  field-reversed configuration with an odd-parity rotating magnetic field},
  Phys. Plasmas 9 (2002) 2093--2102.

\bibitem{welch2010formation}
D.~Welch, S.~A. Cohen, et~al., Formation of field-reversed-configuration plasma
  with punctuated-betatron-orbit electrons, Phys. Rev. Lett. 105 (2010) 015002.

\bibitem{steinhauer2011review}
L.~C. Steinhauer, Review of field-reversed configurations, Phys. Plasmas 18
  (2011) 070501.

\bibitem{ishida1988variational}
A.~Ishida, H.~Momota, L.~Steinhauer, Variational formulation for a multifluid
  flowing plasma with application to the internal tilt mode of a field-reversed
  configuration, Phys. fluids 31 (1988) 3024.

\bibitem{dolan2013magnetic}
T.~J. Dolan, R.~W. Moir, W.~Manheimer, L.~C. Cadwallader, Magnetic fusion
  technology, Springer, London, 2013.

\bibitem{zylstra2017proton}
A.~Zylstra, J.~Frenje, et~al., Proton spectra from $^3${He}+t and
  $^3${He}+$^3${He} fusion at low center-of-mass energy, with potential
  implications for solar fusion cross sections, Phys. Rev. Lett. 119 (2017)
  222701.

\bibitem{casey2012measurements}
D.~Casey, J.~Frenje, M.~G. Johnson, et~al., Measurements of the t(t,2n)$^4${He}
  neutron spectrum at low reactant energies from inertial confinement
  implosions, Phys. Rev. Lett. 109 (2012) 025003.

\bibitem{sayre2013measurement}
D.~B. Sayre, C.~R. Brune, et~al., Measurement of the t+t neutron spectrum using
  the national ignition facility, Phys. Rev. Lett. 111 (2013) 052501.

\bibitem{harrison1967radiative}
W.~Harrison, W.~Stephens, T.~Tombrello, H.~Winkler, Radiative capture of
  $^3${He} by $^3${He}, Phys. Rev. 160 (1967) 752.

\bibitem{khvesyuk1995ash}
V.~I. Khvesyuk, N.~V. Shabrov, A.~N. Lyakhov, Ash pumping from mirror and
  toroidal magnetic confinement systems, Fusion Technol. 27 (1995) 406.

\bibitem{sawan2002impact}
M.~Sawan, S.~Zinkle, J.~Sheffield, {Impact of tritium removal and $^3${He}
  recycling on structure damage parameters in a D--D fusion system}, Fusion
  Eng. Des. 61 (2002) 561.

\bibitem{cohen2015reducing}
S.~Cohen, M.~Chu-Cheong, R.~Feder, et~al., Reducing neutron emission from small
  fusion rocket engines, in: 66th International Astronautical Congress, IAC
  2015, Jerusalem, Israel.

\bibitem{paluszek2014direct}
M.~Paluszek, G.~Pajer, et~al., {Direct fusion drive for a human Mars orbital
  mission}, in: 65th International Astronautical Congress, IAC 2014, Toronto,
  Canada.

\bibitem{turner2003m}
G.~Turner, et~al., Noble gas geochemistry. cambridge (cambridge university
  press), Mineral Mag. 67 (2003) 418.

\bibitem{kennedy2018interstellar}
R.~Kennedy, The interstellar fusion fuel resource base of our solar system, J.
  Br. Interplanet. Soc. 71 (2018) 298--305.

\bibitem{fa2010global}
W.~Fa, Y.~Jin, Global inventory of helium-3 in lunar regoliths estimated by a
  multi-channel microwave radiometer on the {Chang-E} 1 lunar satellite, Sci.
  Bull. 55 (2010) 4005--4009.

\bibitem{wittenberg1986lunar}
L.~Wittenberg, J.~Santarius, G.~Kulcinski, Lunar source of $^3${He} for
  commercial fusion power, Fusion technol. 10 (1986) 167--178.

\bibitem{fa2007quantitative}
W.~Fa, Y.-Q. Jin, Quantitative estimation of helium-3 spatial distribution in
  the lunar regolith layer, Icarus 190 (2007) 15--23.

\bibitem{mcgreivy2016}
N.~McGreivy, {SULI Report: UEDGE simulations of a direct fusion drive FRC
  rocket}, Tech. rep., Princeton Plasma Physics Laboratory (2016).

\bibitem{vinti1973gaussian}
J.~P. Vinti, Gaussian variational equations for osculating elements of an
  arbitrary separable reference orbit, Celest. Mech. 7 (1973) 367.

\bibitem{JPLHorizons}
{NASA-JPL Solar System Dynamics}, {JPL HORIZONS on-line solar system data and
  ephemeris computation service}, \url{{https://ssd.jpl.nasa.gov/?horizons}}
  (2019).

\bibitem{dvornychenko1990generalized}
V.~Dvornychenko, The generalized tsiolkovsky equation, in: NASA Conference
  Publication, Vol. 3102, National Aeronautics and Space Administration,
  Scientific and Technical Report, 1990, p. 449.

\bibitem{kosmodemyansky2000konstantin}
A.~Kosmodemyansky, X.~Danko, Konstantin Tsiolkovsky: His Life and Work, The
  Minerva Group, Inc., 2000.

\bibitem{boltz1992orbital}
F.~W. Boltz, Orbital motion under continuous tangential thrust, J. Guid.
  Control Dyn 15 (1992) 1503--1507.

\bibitem{bate2019fundamentals}
R.~R. Bate, D.~D. Mueller, J.~E. White, W.~W. Saylor, Fundamentals of
  astrodynamics, Dover publications, 2019.

\bibitem{harland2002mission}
D.~M. Harland, Mission to Saturn: Cassini and the Huygens probe, Springer
  Science \& Business Media, 2002.

\bibitem{mitchell2006cassini}
R.~T. Mitchell, Cassini/huygens at saturn and titan, Acta Astronaut. 59 (2006)
  335--343.

\bibitem{NASA}
\url{https://www.jpl.nasa.gov/missions/cassini-huygens/}.

\bibitem{ESA}
\url{https://www.esa.int/Science_Exploration/Space_Science/Cassini-Huygens/Cassini_spacecraft}.

\bibitem{johnson2005power}
K.~S. Johnson, R.~D. Cockfield, Power and propulsion for the cassini mission,
  in: AIP Conference Proceedings, Vol. 746, American Institute of Physics,
  2005, pp. 232--239.

\bibitem{tobie2006episodic}
G.~Tobie, J.~I. Lunine, C.~Sotin, {Episodic outgassing as the origin of
  atmospheric methane on Titan}, Nature 440 (2006) 61--64.

\bibitem{elachi2005cassini}
C.~Elachi, S.~Wall, et~al., {Cassini radar views the surface of Titan}, Science
  308 (2005) 970.

\bibitem{acuna1980magnetic}
M.~H. Acu{\~n}a, N.~F. Ness, The magnetic field of saturn: Pioneer 11
  observations, Science 207 (1980) 444--446.

\bibitem{smith1981encounter}
B.~A. Smith, L.~Soderblom, et~al., Encounter with saturn: Voyager 1 imaging
  science results, Science 212 (1981) 163--191.

\bibitem{smith1982new}
B.~A. Smith, L.~Soderblom, et~al., A new look at the saturn system: The voyager
  2 images, Science 215 (1982) 504--537.

\bibitem{smith2020artemis}
M.~Smith, D.~Craig, et~al., The artemis program: An overview of nasa's
  activities to return humans to the moon, in: 2020 IEEE Aerospace Conference,
  IEEE, 2020, pp. 1--10.

\bibitem{ESAweb}
{The European Space Agency (ESA)}, {N° 6–2020: ExoMars to take off for the
  Red Planet in 2022},
  \url{{https://www.esa.int/Newsroom/Press_Releases/ExoMars_to_take_off_for_the_Red_Planet_in_2022}}
  (2020).

\end{thebibliography}
\end{document}